\documentclass{elsarticle}
\usepackage[margin=3.4cm]{geometry}
\usepackage{latexsym}
\usepackage{longtable}
\usepackage{float}
\usepackage[latin2]{inputenc}
\usepackage{graphicx}
\usepackage{ae}
\usepackage{aecompl}
\usepackage{setspace}
\usepackage{amsmath}
\usepackage{amssymb}
\usepackage{amsthm}
\usepackage[hidelinks]{hyperref}
\usepackage{pdfpages}

\makeatletter
\def\ps@pprintTitle{%
	\let\@oddhead\@empty
	\let\@evenhead\@empty
	\def\@oddfoot{\centerline{\thepage}}%
	\let\@evenfoot\@oddfoot}
\makeatother

\begin{document}

\title{On the Increased and Decreased Connectivity of the Demented Human Brain}

\author[p]{D\'aniel Heged\H{u}s}
\ead{hegedus@pitgroup.org}
\author[p]{M\'arton Barnab\'as M\'ora}
\ead{mora@pitgroup.org}
\author[p]{B\'alint Varga}
\ead{balorkany@pitgroup.org}
\author[p,u]{Vince Grolmusz\corref{cor1}}
\ead{grolmusz@pitgroup.org}
\cortext[cor1]{Corresponding author}
\address[p]{PIT Bioinformatics Group, ELTE E\"otv\"os Lor\'and University, H-1117 Budapest, Hungary}
\address[u]{Uratim Ltd., H-1118 Budapest, Hungary}

\date{}

\begin{abstract}
With major advances in cerebral imaging techniques, a large amount of data is available for studying the aging and demented brain. In this contribution, we apply the OASIS-3 dataset to identify small areas of human gray matter with higher or lower structural connectivity in dementia. As was anticipated, we have found that finer structures of the hippocampus and the temporal lobe have decreased connectivity in dementia. More surprisingly, the precuneus, the cuneus, and some finer structures of the insula, the paracentral lobule, and the precentral and paracentral gyri showed higher connectivity in dementia than in healthy subjects.  
\end{abstract}

\maketitle

\section*{Introduction}

As human life expectancy increases, age-related diseases have become a much greater concern worldwide than several decades ago. Age-related anatomical changes are well-documented in various organs \cite{whitbourne2012aging}, including reductions in the weight and volume of muscles, lungs, kidneys, and the liver. Bone mass decreases, skin becomes thinner and less elastic, and cartilage degeneration leads to painful joint movement. Additionally, arterial walls thicken and harden, leading to cardiovascular aging. Understanding these age-related physiological changes has paved the way for numerous treatments, novel medications, and surgical interventions, all aimed at promoting healthier aging and prolonging independent living.

The human brain and its connections also change during aging. These functional changes are often linked to anatomical aging \cite{Jauny2024}.  Several studies have reported age-associated reductions in cortical volumes \cite{Asken2023} or subcortical gray matter volumes \cite{Bigler1997}, and changes in anatomical brain connections \cite{Zhao2015, Damoiseaux2017, Wang2022,Lin2008}.

One of the most important age-related changes in brain function is dementia. Dementia is uncommon before age $60$, but its prevalence doubles approximately every five years thereafter \cite{Bermejo-Pareja2008, Carlo2002}. It affects around $40\%$ of individuals over $90$ and up to $20\%$ of those between $75$ and $84$ \cite{Wortmann2012, war2009}. According to recent World Health Organization (WHO) estimates \cite{Wortmann2012}, $55$ million people globally have dementia, with $10$ million new cases diagnosed annually. The leading cause of dementia is Alzheimer's disease (AD), which initially manifests as forgetfulness, disorientation, and impairments in concentration, calculation, language, and judgment. As the disease progresses, some patients experience severe behavioral disturbances and psychosis. In its final stages, individuals with AD lose the ability to care for themselves and become bedridden \cite{Wortmann2012, war2009}.

The economic impact of dementia is also staggering, with an estimated yearly cost of $1.3\cdot 10^{12}$ \$ US worldwide (estimated for the year $2019$, \cite{Wimo2023}), including direct medical costs, direct social sector costs, and informal social costs.

In this contribution, we compare local properties of anatomical brain connections in demented and healthy subjects using a large, publicly available dataset from \url{https://braingraph.org} \cite{Varga2024}.

\subsection*{Braingraphs} 

Connectomes, also known as braingraphs, have become important tools in human brain research. Two primary approaches exist for their construction: functional connectomes derived from functional MRI (fMRI) and structural or anatomical connectomes based on diffusion MRI. In this work, we focus exclusively on the latter.

Diffusion MRI enables characterization of macroscopic water-molecule diffusion within neuronal structures. In white matter, water molecules predominantly diffuse along the direction of neuronal axons, allowing for the identification of axonal pathways connecting different regions of gray matter within the cortex and subcortex. 

By labeling these gray matter regions with their anatomical names, a connectome or braingraph can be constructed as follows: the nodes represent anatomically labeled gray matter areas, commonly referred to as Regions of Interest (ROIs), and edges are established between two nodes if a diffusion MRI-based tractography workflow \cite{Besson2014a} identifies an axonal bundle linking them.

This process effectively transforms MRI imaging data into a discrete graph structure, enabling the application of the extensive and well-developed field of mathematics, called graph theory. Originally introduced by Leonhard Euler in $1741$ \cite{Eulera}, graph theory has matured significantly over the 20th and 21st centuries, thanks to contributions from numerous mathematicians \cite{HBComb}.

Our research group has computed and published multiple sets of undirected and directed braingraphs, each containing up to $1015$ vertices per brain \cite{Kerepesi2016b,Szalkai2015a,Szalkai2016,Kerepesi2015b, Szalkai2016d}, derived from various public data releases of the Human Connectome Project \cite{McNab2013}. These datasets, available at \url{https://braingraph.org}, have been widely used in structural analyses of the young and healthy human brain \cite{Szalkai2015,Kerepesi2015a, Szalkai2016d, Kerepesi2016, Szalkai2016c, Szalkai2017c, Szalkai2016e, Szalkai2016a,Fellner2017,Fellner2019,Fellner2018}.

Because the Human Connectome Project releases MRI data for young, healthy individuals, those public datasets do not support the study of the aging brain. Here we examine braingraphs from Section A at the site \url{https://braingraph.org/cms/download-pit-group-connectomes/} \cite{Varga2024}, for studying the aging healthy and demented human brain \cite{Varga2024}. The data source of those graphs is the public release of the OASIS-3 dataset \cite{LaMontagne2019}, which contains MRI and PET records of 1098 subjects aged between 42 and 95 years. The data recordings entail a period of 15 years on 3T Siemens MR scanners at the Washington University Knight Alzheimer Disease Research Center \cite{LaMontagne2019}.

In the present contribution, we evaluate some relevant graph-theoretical differences between the graphs of healthy and demented brains.

In addition to the individual braingraphs, we have constructed averaged graphs for several cohorts, including all subjects, healthy and demented subjects, males, females, healthy males, healthy females, demented males, and demented females. These averaged graphs are publicly available along with the individual connectomes and can be used for group-level analyses and comparisons. The graphs can be accessed at \url{https://braingraph.org/cms/download-pit-group-connectomes/} under sections A and A1.

\subsection*{Braingraphs and dementia} 

Today, dementia is most often diagnosed with the help of clinical syndromes and cognitive tests \cite{Furtner2021}. Medical imaging, mostly MRI, is applied in auxiliary roles to rule out brain tumors or hematomas, and to evaluate possible vascular damage, microbleeds or ischemia, as well as atrophy of cerebral regions \cite{Furtner2021}. 

Although graph-theoretical changes are not yet used to diagnose dementia in clinical practice, many studies in the literature address them, for example  \cite{Daianu2013,Ibanez2014,Li2013e,Wang2013c,Xia2014,Lazarou2020,Yu2024,Xu2024,Chong2024}. Most of these works analyze connections between braingraphs and dementia, particularly Alzheimer's disease, using network science tools and terminology.  In this work, we compare graph-theoretical properties of healthy and demented braingraphs.

\section*{Methods}

In the present work, we analyze the braingraphs published in Section A at \url{https://braingraph.org/cms/download-pit-group-connectomes/} \cite{Varga2024}, computed by our research group.

\subsection*{Braingraph construction from the OASIS-3 dataset}

 The data source is the public release of the OASIS-3 dataset \cite{LaMontagne2019}. The resource contains MRI and PET imaging recordings together with rich clinical data, from 1098 subjects between the ages 42 and 95 years, over a 15-year time span.  Diffusion MRI data from 1472 sessions were recorded using Siemens 3T scanners of two models: TIM Trio 3T, and BioGraph mMR PET-MR 3T at the Washington University Knight Alzheimer Disease Research Center \cite{LaMontagne2019}.
 
 \subsection*{Computational Workflow}
 
 The computation of braingraphs involved the identification of anatomically labeled gray matter areas (parcellation) and the computation of axonal fiber tracts (or streamlines) which connect those gray matter areas (tractography). The resulting braingraph has a vertex set corresponding to the anatomically labeled gray matter areas, and two nodes are connected by an edge if tractography finds at least one axonal fiber connecting them. The gray matter areas were labelled by the FreeSurfer tool.
 
 For the computation, we have applied the Connectome Mapper Tool Kit v.3.1., abbreviated CMP3.1 \cite{Tourbier2022,TourbierZenodo}, with probabilistic tractography. CMP3.1 was applied with diffusion spectrum imaging (DSI) modeling, with one million streamlines.  The most relevant MRtrix3 tractography parameters were  {\tt mrtrix\_tracking\_config.min\_length= 5.0, mrtrix\_tracking\_config.max\_length=500.0, mrtrix\_tracking\_config.angle=45.0, mrtrix\_tracking\_config.cutoff\_value=0.05.}

 From each MRI dataset, we have prepared 5 graphs with different resolutions: 124, 170, 272, 502 and 1058 vertices, using the FreeSurfer segmentation tool, according to the Lausanne2018 brain parcellations \cite{Tourbier2022,TourbierZenodo}. We have also computed and deposited group-averaged graphs for all subjects, healthy and demented subjects, males, females, healthy males, healthy females, demented males, and demented females.
  
  We processed the data of 696 subjects from OASIS-3; the remaining subjects had one or more missing or erroneous files, which prevented processing with the CMP3.1 workflow. More than one MRI were processed from numerous subjects, therefore, we have computed 975 graphs from the diffusion MRI data: one dataset from 482 subjects, 2 datasets from 156 subjects, 3 datasets from 52 subjects, 4 datasets from five subjects, 5 datasets from 1 subject. In the present study, we have applied the most detailed 1058-vertex graphs from the repository \url{https://braingraph.org/cms/download-pit-group-connectomes/} exclusively.

  In the braingraphs deposited at \url{https://braingraph.org}, the nodes and edges have several attributes, detailed in the Appendix. In the present work, we use the anatomical labels of the vertices, and the edge weights describing fiber numbers; that is, the number of axonal tracts identified between the endpoints of the edge by the tractography phase of the processing. This quantity corresponds to the width of the connection, which can also be represented as a strength or the importance of a connection \cite{Hegedus2022, Hegedues2025}.

\subsection*{Assigning ``demented'' and ``healthy'' labels to the graphs}
 
The OASIS-3 dataset \cite{LaMontagne2019} contains psychiatric diagnosis for each imaging session for each subject. Since some subjects were evaluated several times, they have several associated braingraphs, corresponding to multiple recordings. We have assigned the ``healthy'' and ``demented'' labels to the braingraphs in the following way: 

If any of the cognitive diagnoses was ``demented'' for the person, then all of his/her graphs were labeled ``demented''. Otherwise, the graph was labeled ``healthy''.

We note that this way the ``demented'' status also corresponds to pre-dementia, that is, persons who will develop dementia within a several-year interval. Our results, consequently, correspond to anatomical changes in dementia and also in pre-dementia.

In this way, we assigned the ``demented'' label to 351 graphs and the ``healthy'' label to 624 graphs. We did not classify different causes of dementia (e.g., Alzheimer's disease, frontotemporal dementia, etc.).

 \subsection*{Some graph-theoretical notations}

 In this work, we examine the weighted degree of the braingraphs. The weights are assigned to the edges and correspond to the fiber number. 
 
 \subsection*{Weighted degree} 
 
 Suppose that we are given a graph $G$ with edge-set $E$ and vertex-set $V$. A widely known and applied quantity is the {\it degree} of a vertex $v$, which is defined as the number of edges adjacent to $v$; this non-negative integer is denoted by $d(v)$. If we have a weight function $w$ on the edges, then we can generalize the vertex degree to {\it weighted degree} $d_w(v)$, as the sum of the weights on the edges adjacent to vertex $v$:
 $$ d_w(v)=\sum_{\{v,w\}\in E}w(\{v,w\}).$$
 
 In a sense, if a vertex has large degree, then it has some ``importance'', since it is connected to many other vertices. If a vertex has a large weighted degree with a non-negative weight function $w$, then it also has some ``importance'', since it is connected to other vertices with high-weighted edges.
 
 We will use the weighted degree for braingraphs with the fiber number weight function, denoted by $fn$. In this setting, the weighted degree $d_{fn}(v)$ of a vertex $v$ is the total number of axonal tracts (i.e., fibers); that connect $v$ to its neighbors. It should be noted that, for a small number of vertices, no connections have been identified. In these cases, we define the weighted degree to be $0$.

\subsection*{Statistical remarks} 

In this work, we are interested in the braingraph vertices, which show a statistically significant and the largest weighted degree change between the healthy and the demented statuses. For each node, we computed the weighted degree, averaged over demented subjects, and the weighted degree averaged over healthy subjects. Table S1 in the Appendix shows the results. For each vertex, we also computed the ratio of these two quantities. If the ratio is less than 1, then the averaged weighted degree of the vertex is smaller in diseased subjects than in the healthy ones. If the ratio is larger than one, then  averaged weighted degree of the vertex is larger in demented subjects. 

The last column of Table S1 contains the p-values, computed by the Student's two-sided t-tests \cite{Wonnacott1972} individually for each row. 

To control for statistical errors in multiple tests, we applied the Bonferroni adjustment \cite{Wonnacott1972} with a target adjusted p-value of 0.05, with $m=1044$ rows (number of tests) in Table S1,; that is, we say that a weighted degree ratio of a vertex significantly differs from 1, if the  Student's t-test value in the last column of Table S1 is less than $4.7\cdot 10^{-5}$. Tables 1 and 2 in the main text contain only significant Bonferroni-adjusted results.

We note that the Bonferroni adjustment is the strictest, most conservative treatment of multiple test errors; therefore, if a ratio is significant under it, than it will be significant under many other adjustment methods (e.g., Holm-Bonferroni \cite{Holm1979} or Benjamini-Hochberg \cite{Benjamini2001} corrections).

\section*{Results and Discussion}

In Table 1, we list 30 brain areas corresponding to graph vertices, whose weighted degrees decreased most strongly and significantly in dementia. As expected from the literature, the hippocampus and several other gray matter areas show significantly lower weights in dementia than in healthy subjects. However, we found (also in line with some, but far fewer, references in the literature) that some connections are stronger in dementia than in healthy persons. It should be noted that, in these cases, the growth of new axons was considered unlikely \cite{Fawcett2019}, but new results show that in certain cases it seems to be possible \cite{Liu2006b,Holland2015,Gage2025}.

We emphasize that, in these tables, we quantitatively computed weighted degree changes for all cortical and subcortical gray matter areas on a large dataset (351 demented, 621 healthy graphs). Naturally, several nodes with the largest changes have been addressed in the literature, but most studies focused on one or two such areas and rarely presented quantitative results on a large dataset.  We also emphasize that increased structural connectivity in dementia is very rarely reported in the literature before the present study.

\subsection*{Vertices with weakened connections in dementia}

\begin{table}[H]
	\includegraphics[width=14.5cm]{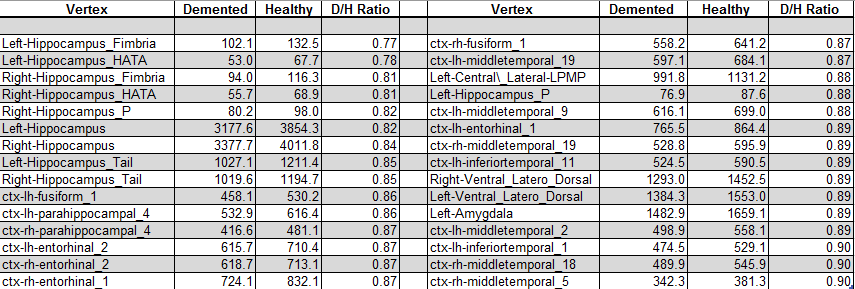}
	\caption{30 vertices with the largest weighted degree decrease in dementia. In the D/H column, the quotient of the averages of the weighted degrees is given in increasing order, taken for the demented (D) and healthy (H) graphs. The Bonferroni-adjusted p-values are 0.05 or less. Abbreviations: Right-Hippocampus\_P=Right-Hippocampus\_Parasubiculum; Left-Hippocampus\_P=Left-Hippocampus\_Parasubiculum; Left-Central\_Lateral-LPMP=Left-Central\_Lateral-Lateral\_Posterior-Medial\_Pulvinar}
\end{table}

Table 1 shows several hippocampal areas with the lowest D/H ratio, meaning we found the largest loss in fiber numbers (represented by edge weights) in those areas. It is well known, that the hippocampus has a definitive role in the short-time memory loss of Alzheimer's disease, and is strongly affected by dementia and AD \cite{Voineskos2015,Nees2014,Bigler1997}. Therefore, our findings for the hippocampus align with the literature. 

In addition to the hippocampus, temporal areas also appear frequently (8 times) in Table 1. In frontotemporal dementia, temporal area shrinkage is well studied \cite{Zhou2014a,Premi2013}. In Alzheimer's disease, memory loss is also associated with temporal areas, and even in early AD, loss of functional connections occurs in the temporal lobes, especially in the middle temporal areas \cite{Grajski2023,Berron2020}.  

The fusiform gyrus appears, in both the right and left hemispheres in Table 1. Volumetric atrophies were reported in the fusiform girus in \cite{Chang2016,Ding2016} and has found to be strongly related to semantic dementia. The work \cite{Cai2015} investigated the functional connectivity changes in amnestic mild cognitive impairment subjects with low-frequency courses in resting-state fMRI, and has found functional connectivity changes involving the fusiform gyrus to some other brain areas, including left precuneus, left lingual gyrus, right thalamus, supramarginal gyrus, left supplementary motor area, left inferior temporal gyrus, and left parahippocampus.

Interestingly, in the first 30 vertices listed in Table 1, the left amygdala is present, while the right is not. This observation aligns with the literature: \cite{Vereecken1994} shows that the neuronal loss is much higher in left amygdala than in the right in dementia; \cite{Cavedo2011} reports similar results in volumetric studies of 38 subjects.

These findings align with earlier literature and validate our methods and results, which present strictly quantified structural connectivity changes in dementia.

\subsection*{Vertices with strengthened connections in dementia}

The data presented in Table 2 is more interesting in several aspects than that of Table 1. Table 2 shows 20 vertices with the largest differences between healthy and demented weighted degrees, where the demented weighted degree is larger. That is, in these areas the connections became stronger in dementia than in the healthy graphs!

\begin{table}[H]
	\includegraphics[width=14.5cm]{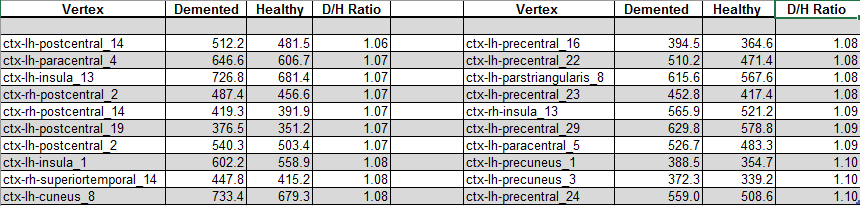}
	\caption{Vertices with the largest weighted degree increase in dementia. We list the 20 vertices with the largest increase. In the D/H column, we report the ratio of the average weighted degrees for the demented (D) and healthy (H) graphs. All Bonferroni-adjusted p-values are 0.05 or less.}
\end{table}

Although we are not aware of previous structural studies showing increased connectivity in dementia, numerous studies have reported increased functional connectivity. As an early result, \cite{Kenny2013} has shown increased functional connectivity in several dementias by fMRI studies on a small cohort. 

Subdivisions of the left precuneus are present twice in Table 2. In \cite{Aponte2025}, the precuneus showed increased functional connectivity in Alzheimer's disease. Here, we have found, that precuneus have increased structural connectivity in dementia, as well.  In \cite{Fischer2025}, greater functional activity of the precuneus is measured in early Alzheimer's disease. Now we have shown higher structural connectivity in our structural data.

Edges connected to the insular cortex also become stronger in dementia, as shown in Table 2; similar findings for the increased gyrification of the insula were reported in \cite{Coleman2025}. Increased compensatory functional connectivity was found in \cite{Zheng2025} in insula subdivisions. \cite{Liu2018} found some insular connections with increased functional connectivity, while others showed decreased functional connectivity. We have shown that, on average, demented subjects have higher structural connectivity to or from the insula.

The precentral gyrus, the paracentral lobule and the postcentral gyrus are also frequently appeared in Table 2, while it is not yet shown that these areas have stronger functional connectivity in dementia.

\section*{Conclusions} We have examined the structural connections originating or ending in specific brain areas in healthy and demented subjects in the OASIS-3 dataset. 1058 brain areas were considered, from which 1044 contained adjoining connections. The cohort contained 696 subjects, and 975 braingraphs, since some of the subjects were recorded multiple times. We assigned the "demented" label to all graphs of a given subject if, at any examination, the "demented" diagnosis was given. This way, we studied 351 graphs labeled "demented" and 624 graphs labeled "healthy". 
We aimed to identify the brain areas most strongly affected by connection changes in dementia. Table 1 lists areas related to decreased structural connectivity, and Table 2 lists areas related to increased structural connectivity in dementia.

\section*{Acknowledgments}
Data were provided in part by OASIS-3 Longitudinal Multimodal Neuroimaging: Principal Investigators: T. Benzinger, D. Marcus, J. Morris; NIH P50 AG00561, P30 NS09857781, P01 AG026276, P01 AG003991, R01 AG043434, UL1 TR000448, R01 EB009352. DH, MBM, BV and VG were partially supported by the ELTE TKP 2021-NKTA-62 project. DH was partially supported by The EK\"OP-25 University Research Scholarship Program of the Ministry for Culture and Innovation from the source of The National Research, Development and Innovation Fund.

\bigskip

\section*{Data availability} The braingraphs are available under Section A at the site \url{https://braingraph.org/cms/download-pit-group-connectomes/}. This study used the Scale 5 - 1058 nodes set exclusively. The website also contains group-averaged graphs according to sex and health status under section A1.

\medskip

\noindent Conflict of Interest: The authors declare no conflicts of interest.




\section*{Appendix}

Here we describe the node- and edge attributes in the GraphML files published in \url{https://braingraph.org/cms/download-pit-group-connectomes/}.

 \subsection*{Node attributes}

The node attributes in the GraphML files include the following values:
\begin{itemize}
	\item dn\_region can be cortical or subcortical
	
	\item dn\_position\_x, dn\_position\_y, dn\_position\_z the coordinates of a vertex. In a few cases, mostly in the substructures of the hippocampus, their values are missing if the substructure is not identified reliably. In that case, the NAN abbreviation (not a number) is given there.
	
	\item dn\_name the corresponding anatomical name; it is either identical to or a refinement of the fsname attribute.
	
	\item dn\_hemisphere  left or right 
	
	\item dn\_fsname  the corresponding anatomical area name of the node in FreeSurfer.
	
	\item dn\_multiscaleID  numerical node ID 
	
	When, in lower resolutions, the sub-partitioning of small areas, mostly subcortical structures, cannot be distinguished from each other (e.g., the finer partitioning of the hippocampus), then the coordinate fields contain the word "nan".
	
\end{itemize}

\subsection*{Edge attributes}

The edge attributes in the GraphML files include the following quantities:

\begin{itemize}
	\item number\_of\_fibers, corresponding to an edge. 
	
	\item fiber\_length mean, median, and standard deviation (std) of the fiber lengths in mm, corresponding to an edge. 
	
\item fiber\_density, normalized\_fiber\_density, fiber\_proportion:  Since fibers (or streamlines) can not always be tracked reliably in tractography algorithms, and since fibers may start or end erroneously in white matter during tractography (which is possible only in gray matter anatomically), some authors prefer to use fiber density quantities instead of or besides fiber numbers \cite{Zhang2022}.
	\item 
	
	The following quantities are related to the image reconstruction method SHORE: Simple harmonic oscillator based reconstruction and estimation \cite{Ozarslan2008, Koay2012,Ozarslan2013}:
	
	\item shore\_rtop\_signal: RTOP: Return-to-Origin Probability \cite{Descoteaux2011}.
	
	\item shore\_msd: MSD: Mean Squared Displacement, with standard deviation (std), mean and median values; \cite{Wu2007, Wu2008b}
	
	\item shore\_gfa: GFA:  derived Generalized Fractional Anisotropy (GFA) with standard deviation (std), mean and median values \cite{Tuch2004};
	
\end{itemize}

\subsection*{Demented-healthy change for all vertices}

The following Table S1 contains the averaged fiber-number-weighted vertex degree for healthy and demented subjects. In the computation, we used graphs from 351 demented and 624 healthy subjects, where ``demented'' includes graphs with corresponding subjects who were not diagnosed as ``demented'' at the time of MRI recording but were later diagnosed in later sessions. The table contains 1044 rows, ordered by increasing D/H ratio. We considered 1058-vertex graphs; the missing 14 rows correspond to 14 vertices that were not connected to any edges, so we did not list those zero-degree vertices here. The p-values indicated here are not multiple-test corrected (for multiple-test correction, refer to the "Statistical remarks " subsection in the "Methods" section).

\begin{longtable}[htbp]{lrrrr}
	\centering

    Name  & \multicolumn{1}{c}{Demented} & \multicolumn{1}{c}{Healthy} & \multicolumn{1}{c}{D/H Ratio} & \multicolumn{1}{c}{p-value} \\
          &       &       &       &  \\
    Left-Hippocampus\_Fimbria & 102.1 & 132.5 & 0.77  & 2.64E-21 \\
    Left-Hippocampus\_HATA & 53.0  & 67.7  & 0.78  & 2.51E-15 \\
    Right-Hippocampus\_Fimbria & 94.0  & 116.3 & 0.81  & 1.91E-14 \\
    Right-Hippocampus\_HATA & 55.7  & 68.9  & 0.81  & 1.98E-12 \\
    Right-Hippocampus\_Parasubiculum & 80.2  & 98.0  & 0.82  & 1.72E-10 \\
    Left-Hippocampus & 3177.6 & 3854.3 & 0.82  & 3.08E-31 \\
\end{longtable}%

\end{document}